# LABORATORY TEST RESULTS UNDERWATER NAVIGATION SYSTEM FOR ENVIRONMENTAL CONTROL DEVICES

A.N. Grekov, M.A Pasynkov, . S.S. Peliushenko

Institute of Natural-Technical Systems
RF, Sevastopol, st. Lenina, 28
E-mail: oceanmhi@ya.ru

The work analyzes existing methods for increasing the accuracy of determining weekend navigation parameters of unmanned underwater vehicles. It is proposed to retrofit the navigation system with an additional hydrodynamic tilt unit, which will increase accuracy of coordinate determination. A prototype of the proposed system was made, and algorithmic software for it. Analysis of typical MEMS errors has been completed accelerometers and gyroscopes. To estimate stochastic errors, the Allan variation method was tested using a set of real data obtained on a prototype as input parameters. A model of stochastic errors of all measuring channels has been constructed

Introduction. An urgent task facing developers of unmanned underwater vehicles (UUVs) (autonomous uninhabited underwater vehicles, remotely controlled uninhabited underwater vehicles), – increase accuracy of determining output navigation parameters: orientation angles, linear velocities and coordinates locations. Definition systems course (eng. AHRS - Attitude and Heading Reference Systems) and spatial provisions that imply equip UUVs, should not be expensive, but technologically suitable for mass production at an acceptable accuracy of coordinate determination. Such devices include strapdown inertial navigation systems (INS) with electronic compass and pressure sensor. An integral part of the INS is a block of sensitive elements consisting of three orthogonally located gyroscopes and three orthogonally located accelerometers. The basis of electronic compass is a three-axis magnetometer. However, such systems, with their low cost, have a large error in determining coordinates, namely, due to zero instability, linear drift, output quantization noise signal, random walk angular speed, acceleration, speed and angle. Although sensors based on MEMS technology have many advantages in terms of cost and size, as a rule, their accuracy is insufficient for the implementation of an ANN. Typical accuracy of pitch and roll calculations such ANN in static conditions is about 10 mrad and 20 mrad when calculated heading angle [1]. In the case when the BPA maneuvers, accuracy may decrease by two orders of magnitude due to the difficulty of determining the Earth's gravity vector as a result of vehicle accelerations. To improve MEMS-based ANN, it was Several calibration and filtering methods have been proposed, including the study of a noise component model sensor New metrological assessment methods consisting of calibration linearity and wavelet processing, improve the accuracy and performance of the inertial module MEMS sensor [2]. To correct the orientation of the UUV, it is proposed to use measurement data (hydrostatic pressure) depth, which allows reduce drift and initial offset MEMS sensors. Depth measuring methods and tools are more accurate and stable than ANNs and magnetometers, which reduces error determining the orientation of the entire system [3]. To improve orientation accuracy, data is combined into based on filter adjustment methods (extended Kalman filter and method Adams). The algorithm is reliable to estimate orientation and depth [4]. A possible solution to the problem is the use of GPS, IMU MEMS and pressure sensors for depth measurement in conjunction with DVL (Doppler Velocity Log). Then, using the extended Kalman filter and fusion technique, the data returns the location of the vehicle with an error of several meters after an hour-long dive [5–7]. However, this method cannot restore exactly the original state. Algorithm useful for restoration of the initial state, – is a UKF (sigma point Kalman filter) that examines the nonlinear model for determining initial orientation, MEMS drift and displacement [8– 10]. To improve the positioning accuracy of the system presented in [11], two sets of inertial measurement sensors were used. However, the use of all of the above methods does not provide acceptable UUV positioning accuracy

Main part. The novelty of the proposed development, which is confirmed patent, is that the navigation system is additionally equipped with a hydrodynamic unit tilt (GBN), containing three differential pressure sensors [12], and, with taking into account their work, AHRS behavior models have been created, this will significantly increase the accuracy of determining angles coordinates, which in turn will increase accuracy of determining linear coordinates. The block diagram of a strapdown navigation system with a HDN unit is shown in Fig. 1.

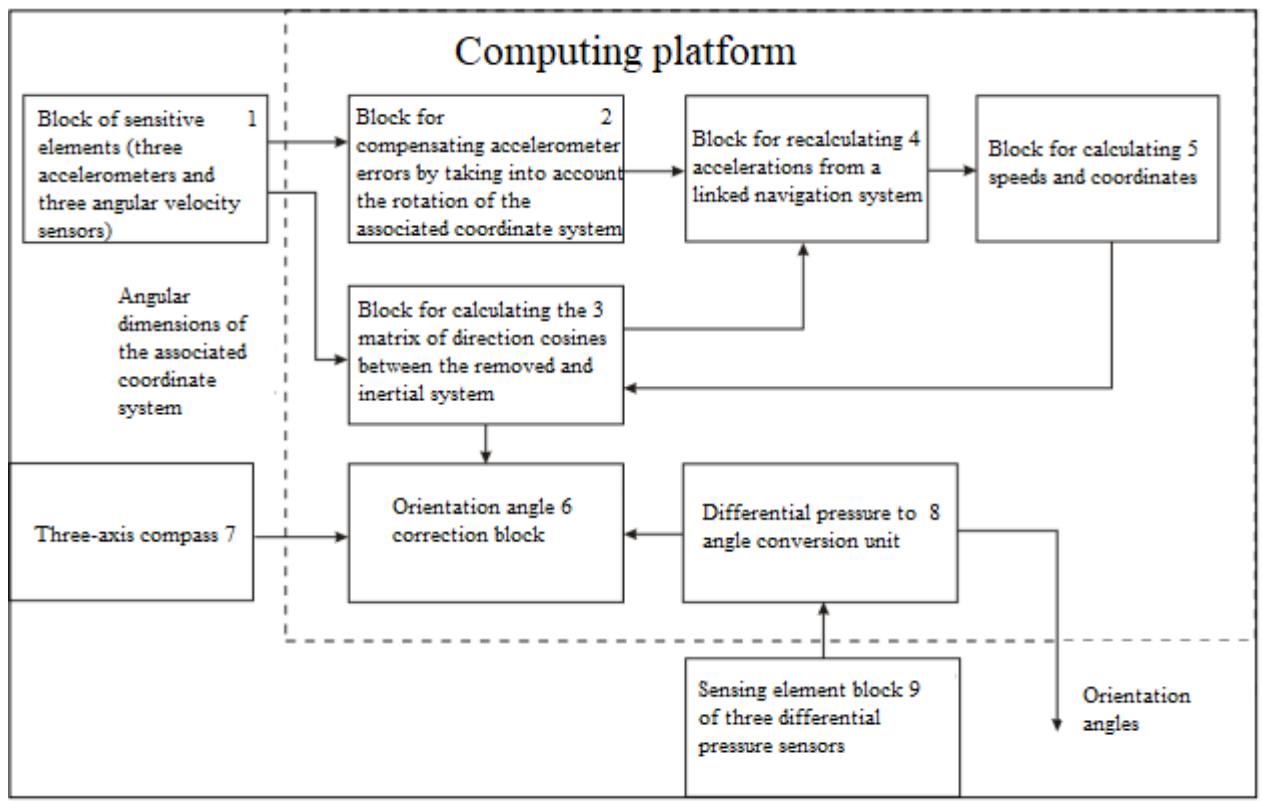

Fig. 1. Structural diagram of a strapdown navigation system

All gyroscopes and accelerometers are subject to errors that limit the accuracy of their measurement of rotation angles and specific forces. Detailed information about sensor errors

$$\omega(t) = M_\omega\big(SF_\omega(t)\omega + B_\omega(t)\big) + W_\omega(t) \quad (1)$$

$$f(t) = M_f\big(SF_f(t)f(t) + B_f(t)\big) + W_f(t) \quad (2)$$

Where $M_{\omega,f}$ are skew–symmetric misalignment matrices; $SF_f$ – scalecoefficients;

w(t) and f(t) represent, respectively, real measurements of gyroscopes and accelerometers;

w(t) and f(t) are ideal measurements; $B_{\omega,f}$ displacement vectors and $W_{\omega,f}$ are white noise vectors. General description $M_{\omega,f}$ and $SF_{\omega,f}$ are given in [13, 15]. Separating all error components may require special settings (for example, tilt-rotary tables), which complicates the measurement procedure. In the case of using MEMS sensors, this may not be necessary, since some error components predominate over other less significant ones, and therefore the error model can be simplified [16–19]. For this reason, the following parameters were ignored: $M_{\omega,f}$, $S_{w,f}^{RC} S_{w,f}^{GM}(t) S_{w,f}^{RW}(t)$

Offsets $B_{\omega,f}(t)$ can be approximated

$$B_\omega(t) = B_\omega^C + B_\omega^{RC} + B_f^{GM}(t) + B_\omega^{RW}(t) + B_\omega^{Bl}(t) \qquad (3)$$

$$B_f(t) = B_f^C + B_f^{RC} + B_f^{GM}(t) + B_f^{RW}(t) + B_f^{Bi}(t) \qquad (4)$$

Where $B_{\omega,c}^C$ constant components displacements (linear drift); stochastic random constants $B_{\omega,f}^{RC}$ processes (random walk of the angle (speeds)), having the distribution:

$$B_{\omega,f}^{GM}(t) = \beta \cdot B_{\omega,f}^{GM}(t) + \omega_k(t)$$

with initial conditions $\lim_{t_0 \to \infty} B_{\omega f}^{GM}(t_0) = 0$ where $\beta = \frac{1}{t}$ is reciprocal of the correlation time, $\omega_k(t)$ is a process Gaussian white noise with zero average $B_{\omega,f}^{RW}(t)$ – random walks of stochastic with initial conditions $B_{\omega,f}^{RW}(t_0) = (B_{\omega,f}^{RW})_0 \cdot B_{w,f}^{BI}(t)$ zero instability that were added to

How:

$$B_{\omega,f}^{BI}(t) = \begin{cases} \sigma_{BI_{w,f}} w_k(t) \text{если } mod(t, T_{BI}) = 0 \\ B_{w,f}^{BI}(t-1) \text{иначе} \end{cases}$$

with an initial condition:
$$B_{w,f}^{BI}(t_0) = \sigma_{BI_{w,f}}, w_k(e_0)$$

Errors in gyroscope sensors and accelerometer in strapdown inertial navigation system can be divided into two parts: the constant (or deterministic) part and stochastic (or random) part. The deterministic part includes itself an offset and a scale factor that can be determined calibration and therefore can be removed from the raw measurements. The random part includes, e.g. drift drift, offset axes and

$B_{\omega,f}^{RC} \sim N(\mu_{\omega,f} \sigma_{\omega,f})$
$B_{\omega,f}^{GM}(t)$ – a random Gaussian Markov process of the first order, defined as:

processes (random walk of angular velocity (acceleration)), defined as:

$$B_{w,f}^{RW}(t) = \omega_k$$

original model developed in [15, 20], in order to more accurately describe the observed noise characteristics. They can be recorded

White noise process (quantization noise
$$W_{w,f}(t) = \sigma_{W_{w,f}}, w_k(t)$$
where $\sigma_{W_{\omega,f}}$ standard deviations of these processes.

random noise. These errors can be modeled in a stochastic model and then included in the vector Kalman filter states. To estimate stochastic errors Allan variation method was applied [21] and tested on an experimental set data obtained for 23 hours on a prototype platform, external the type and functional diagram of which are presented in Fig. 2. Hardware Implementation Components The following systems were chosen:– as a central module a development board was used STM32F401, which provides high speed sensor polling and trajectory calculations; – to determine the position in space, the combined sensor MPU-9250 was used – the sensor of the second generation of InvenSense, which includes a three-axis gyroscope, three-axis accelerometer and three-axis magnetometer; – ADS1115 ADC – 16-bit analog-to-digital converter with 4 exits; – three integrated high-precision piezoresistive differential pressure sensors MPXV7002D.P with temperature compensation and calibration, and with installed liquid compensation protection. To carry out the research there we ralgorithm and software developed provision of the STM32F401 controller, performing setup and survey of all measuring channels, as well as transferring information to a PC via a standard USB interface. Structural scheme

The program is shown in Fig. 3. Preliminary tests of prototype units of underwater inertial navigation system showed that error in determining measured angles, using HDN, did not exceed ~3° for 8 hours of operation.

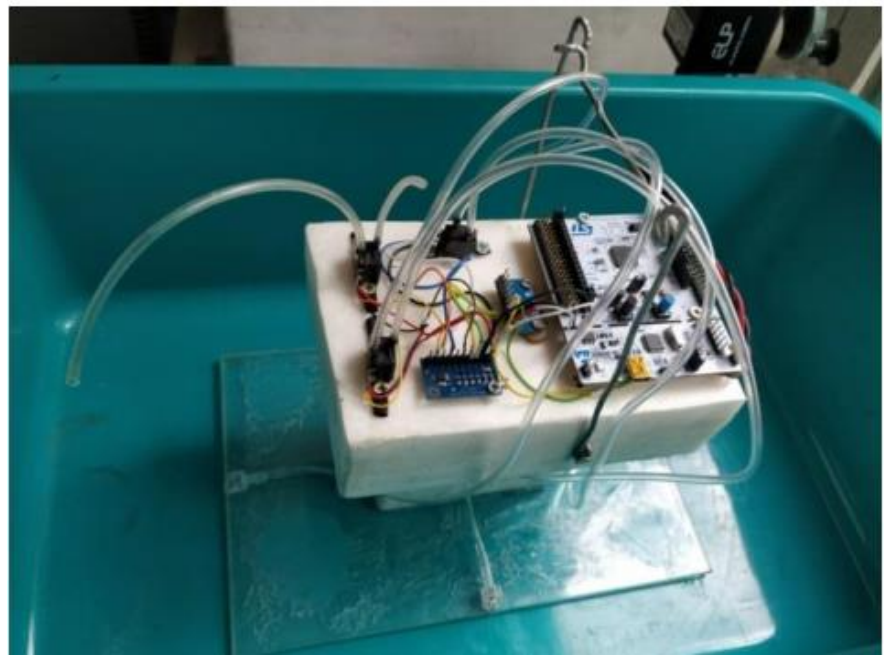

a

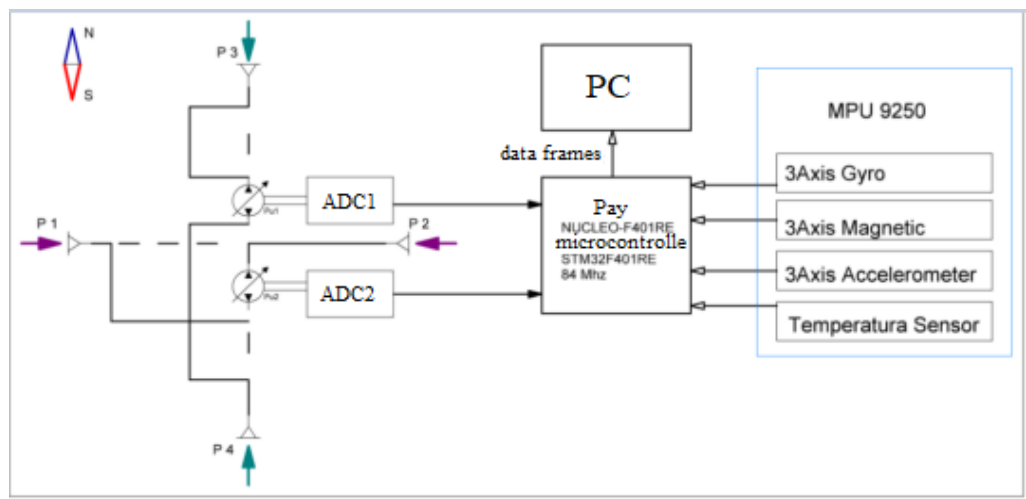

b

Fig. 2. Appearance (a) and functional diagram (b) of the platform layout for research underwater inertial navigation system with MEMS-AHRS modules and hydrostatic sensors

As a result of the experiment, about 2.8 million samples were obtained for each measuring channel of MEMS sensors. Data set obtained in during the 23-hour experiment is shown in Fig. 4. This is an expression of solving only one axis of each measuring channel of the MEMS sensor. The presented graphs characterize stable operation of individual elements of isolated and independent parts of the ANN platform, in particular positive channels and correct implementation of the developed software controller software STM32F401, which is confirmed by the function rationing the system as a whole without it failures during the experiment.

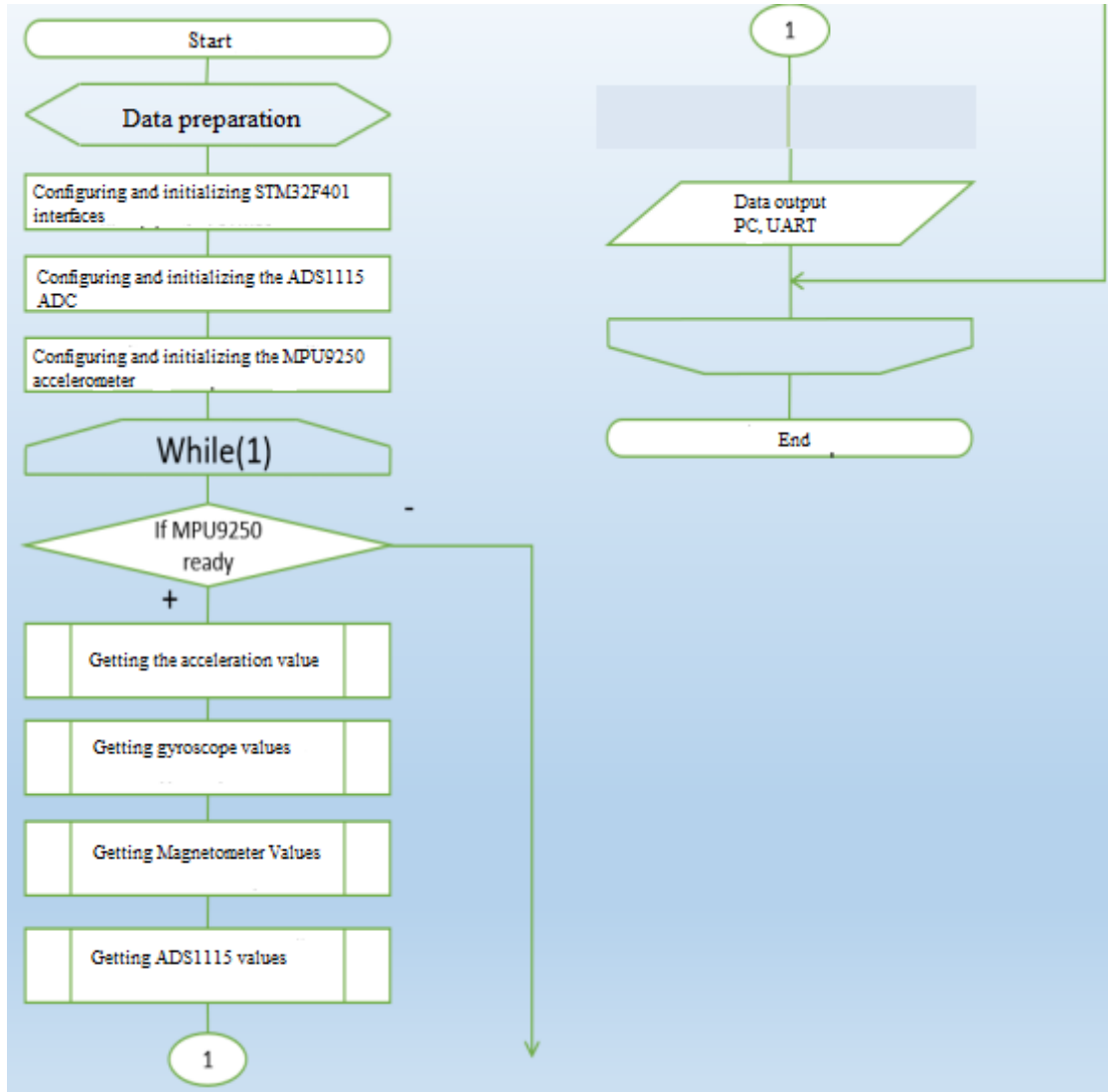

Fig. 3. Block diagram of the developed software for STM32F401 controller

Based on the data obtained from each measuring channel of microelectromechanical sensors, it was calculated Allan variation, the graphical results of which are presented in Fig. 5. The processing results confirm that the presented approach is essentially allows you to include the Allan Variation in layout of measuring navigation system that reflects change errors in time, and its main components provide information about the components of the errors. In our

In this case, this is based on an analysis of basic information about the hardware components implementation of the system and their physical the principles on which they are based. Then, using the method in [22], we calculated all types of errors mentioned earlier (Table 1).

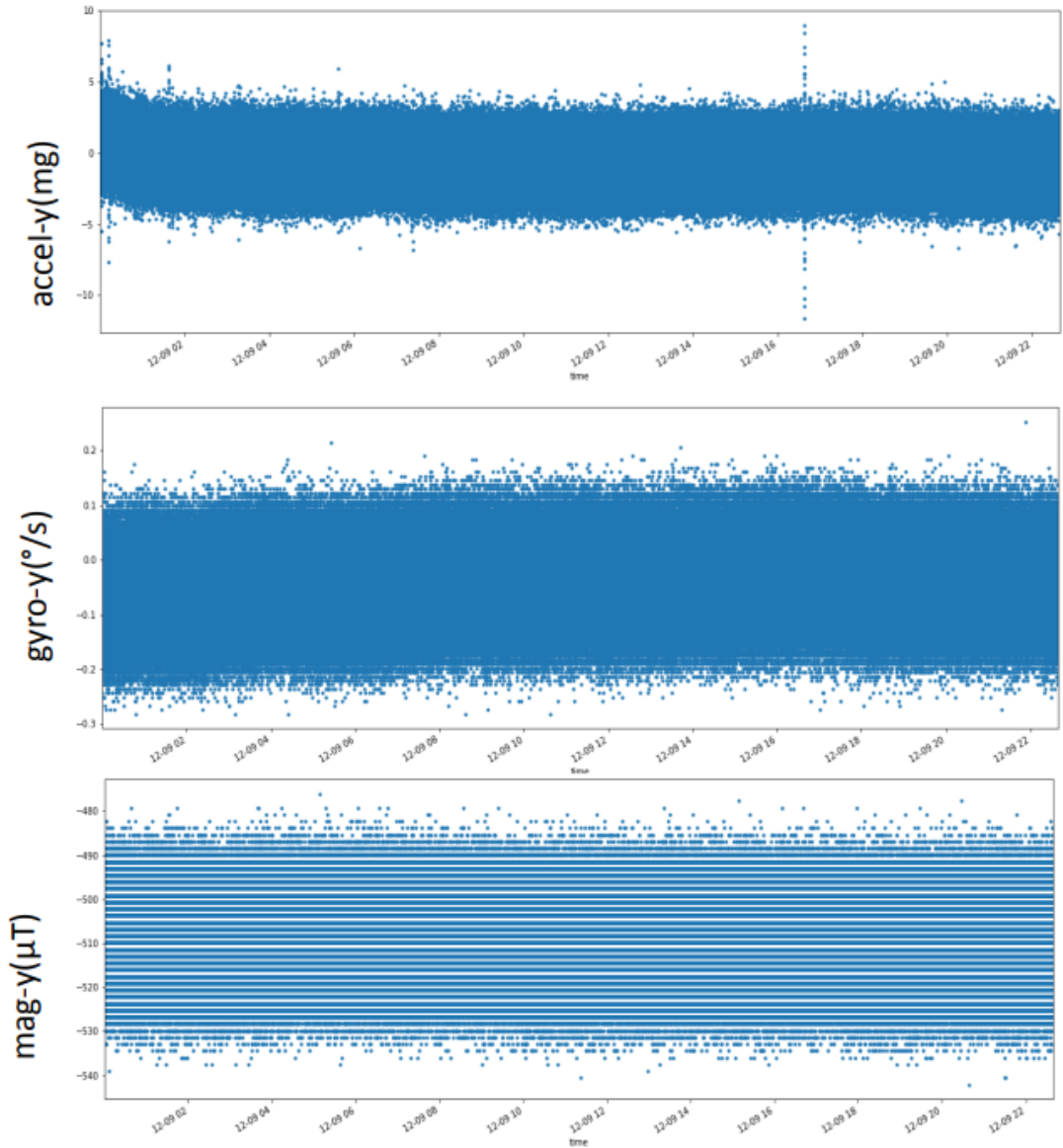

Fig. 4. Measured data set from a 23-hour experiment
(represented by only one measuring axis of each sensor channel MEMS))

To validate the obtained estimates a model of errors of all measuring channels was built, the calculation results of which are presented graphically in Fig. 6.

From the graphs it follows that variations Allan and the error model are in good agreement with each other and these results can be used to form Kalman filter in the resulting system. It should be noted that Allan's variation methods are currently developing most dynamically, and this produces specific variants of their use for estimating stochastic errors in unmanned underwater devices, including other adjacent areas, for example, Marin environmental instrumentation.

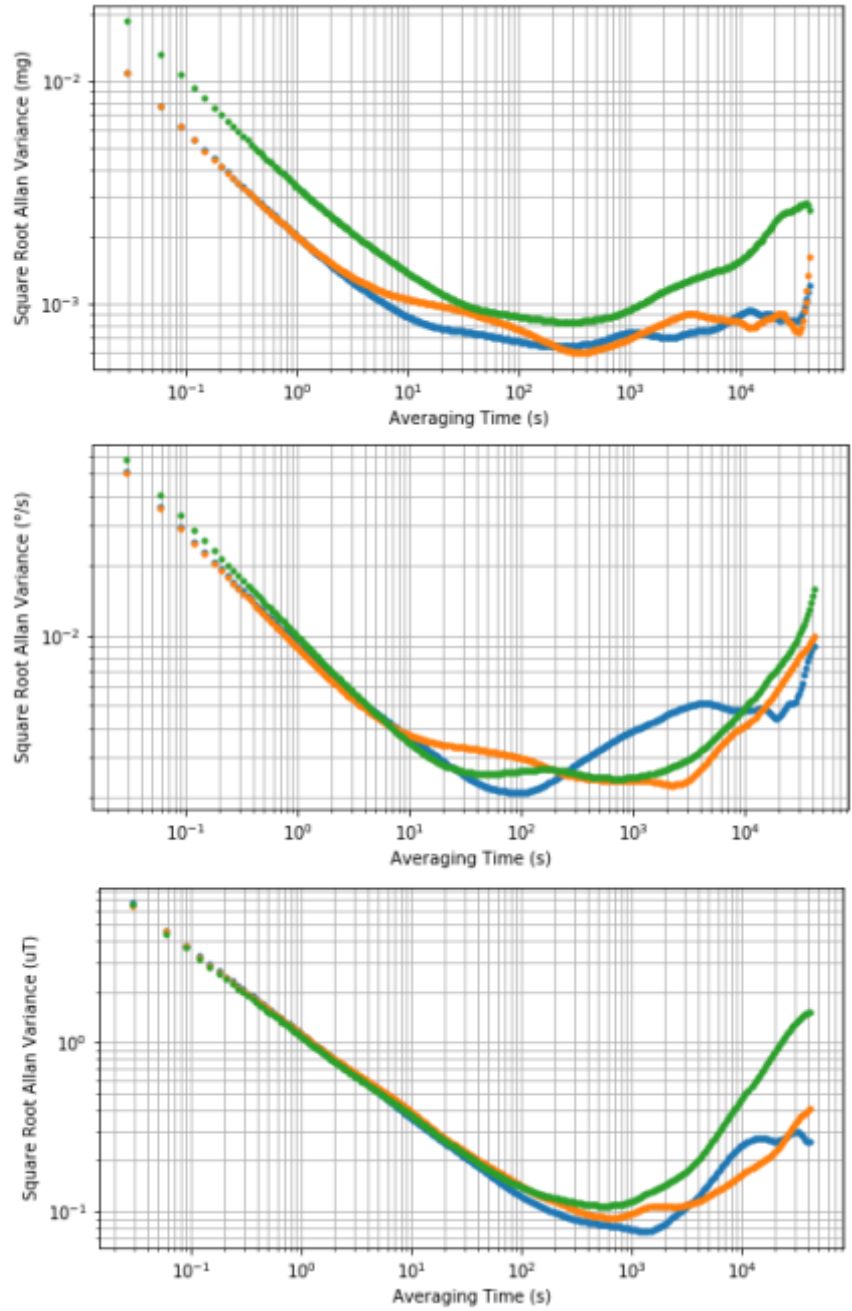

Fig. 5. Allan variance calculation results for each measuring channel
(blue line – X-axis; orange line – Y-axis; green line – Z-axis)

Table 1. Calculation results and types of errors

| Error type | Accelerometer, mg | | | Gyroscope % | | | Magnetometer µT | | |
|---|---|---|---|---|---|---|---|---|---|
| | X | Y | Z | X | Y | Z | X | Y | Z |
| Quantization noise | 0.0 | 0.0 | 0.0 | 0.0002 | 0.0 | 0.0004 | 0.0204 | 0.0 | 0.0093 |
| Random walk of angle (velocity) | 0.0019 | 0.0019 | 0.0032 | 0.006 | 0.086 | 0.0092 | 1.1184 | 1.1370 | 1.0896 |
| Zero speed instability | 0.0010 | 0.0011 | 0.0012 | 0.0034 | 0.0034 | 0.0034 | 0.0749 | 0.1194 | 0.1325 |
| Random walk of angular velocity (acceleration) | 6.7203 | 5.2829 | 2.4941 | 6.1290 | 1.3782 | 5.4810 | 0.0029 | 0.0021 | 0.0027 |
| Multiplicative error (linear drift) | 0.0 | 0.0 | 0.0 | 0.0 | 3.6565 | 4.0463 | 0.0 | 1.0277 | 5.7807 |

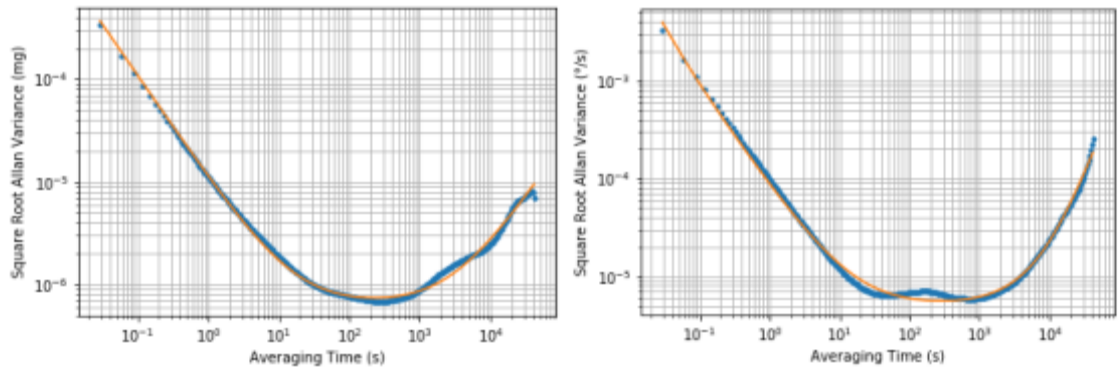

Fig. 6. Calculated Allan variance
(blue dots) and the error model (orange line)

Conclusion. At the first stage of research, a prototype of the navigation platform was developed and manufactured with software for it, on which they were obtained in laboratory experimental data with measuring channels, and then their pre-processing was carried out.

An analysis of existing methods for increasing the accuracy of determination output navigation parameters unmanned underwater vehicles showed that, despite their dynamic development and continuous improvement, remains high and unstable error in calculating coordinates when using MEMS sensors. The use of an additional HDN in the ANN will compensate for MEMS errors sensors that require constant corrections during their stable operation, to unfortunately, only for a short period of time. An analysis of typical errors has been carried out MEMS sensors. Deterministic part of the error of these sensors may be eliminated by calibration, and for estimates of stochastic errors we used Allan's variation. The obtained error values will be used in the future for forming the Kalman filter of the resulting system.